\documentclass[a4paper,amsmath,amssymb,floatfix,pra,footinbib,twocolumn,superscriptaddress,showpacs,preprintnumbers]{revtex4}

\usepackage[dvips]{graphicx}
\usepackage{dcolumn}% Align table columns on decimal point
\usepackage{bm}% bold math
\usepackage{dsfont}
\usepackage{color}

\usepackage{url}
\usepackage[colorlinks=true ,citecolor=blue]{hyperref}
\usepackage{amsmath,amssymb}

\newcommand{\ttt}{$\mathcal{T}_3$~}

\newcommand{\ii}{\text{i}}
\newcommand{\ee}{\text{e}}

\newcommand{\comment}[1]{}

\begin{document}
% 37.10.Jk Atoms in optical lattices 
% 03.75.Hh Static properties of condensates; thermodynamical, statistical, and structural properties
% 05.30.Fk Fermion systems and electron gas
% 71.70.Di Landau levels

\title{Massless Dirac-Weyl Fermions in a \ttt Optical Lattice}
\author{D. Bercioux}
\email{dario.bercioux@frias.uni-freiburg.de}
\affiliation{Freiburg Institute for Advanced Studies, Albert-Ludwigs-Universit\"at, D-79104 Freiburg, Germany}
\affiliation{Physikalisches Institut, Albert-Ludwigs-Universit\"at, D-79104 Freiburg, Germany}
\author{D.~F. Urban}
\affiliation{Physikalisches Institut, Albert-Ludwigs-Universit\"at, D-79104 Freiburg, Germany}
\affiliation{Departamento de F\'isica de la Materia Condensada C-XII, Facultad de Ciencias, Universidad Aut\'onoma de Madrid, E-28049, Madrid, Spain}
\author{H. Grabert}
\affiliation{Freiburg Institute for Advanced Studies, Albert-Ludwigs-Universit\"at, D-79104 Freiburg, Germany}
\affiliation{Physikalisches Institut, Albert-Ludwigs-Universit\"at, D-79104 Freiburg, Germany}
\author{W. H\"ausler}
\affiliation{Physikalisches Institut, Albert-Ludwigs-Universit\"at, D-79104 Freiburg, Germany}
\affiliation{Institut f\"ur Physik, Universit\"at Augsburg, D-86135 Augsburg, Germany}

\date{\today}

\begin{abstract}
We propose an experimental setup for  the observation of quasi-relativistic massless Fermions. It is based on a \ttt optical lattice,  realized by three pairs of counter-propagating lasers, filled with fermionic cold atoms. We show that in the long wavelength approximation the \ttt Hamiltonian generalizes the Dirac-Weyl Hamiltonian for the honeycomb lattice, however, with a larger value of the pseudo-spin $S=1$. In addition to the Dirac cones, the spectrum includes a dispersionless branch of localized states  producing a finite jump in the atomic density. Furthermore, implications for the Landau levels are discussed.
\end{abstract}

\maketitle
%
% INTRODUCTION ON COLD-ATOMS AND OPTICAL LATTICES
%
In the past decade, ultra cold atoms have emerged as a
fascinating new area linking quantum optics with solid state
physics~\cite{Bloch:2008p894}. Essentially, these are the only
quantum many-body systems for which the particle interaction is
both rather precisely known and controllable. In particular,
cold atoms confined in optical lattices
(OLs)~\cite{Greiner:2008p1382} often present systems with
crystalline structure in various spatial dimensions $d=1,2,3$ described by textbook models from solid
state physics with tunable parameters.
This implements Feynman's pioneering idea of quantum simulations
using one physical system to investigate another
one~\cite{FEYNMAN:1982}. A celebrated
example~\cite{Greiner:2002} is the optical realization of the
Mott transition, a well-known phenomenon in solid state physics,
describing the transition from a metal to an insulator with increasing interaction strength.
Furthermore, the possibility to realize an effective magnetic field
by rotation of cold atoms in OLs~\cite{jaksch:2003}
has opened up prospects of studying other fundamental phenomena
in a controlled manner such as the fractional quantum Hall
effect in $d=2$~\cite{sorensen:2005}.

\begin{figure}[t]
	\centering
	\includegraphics[width=\columnwidth]{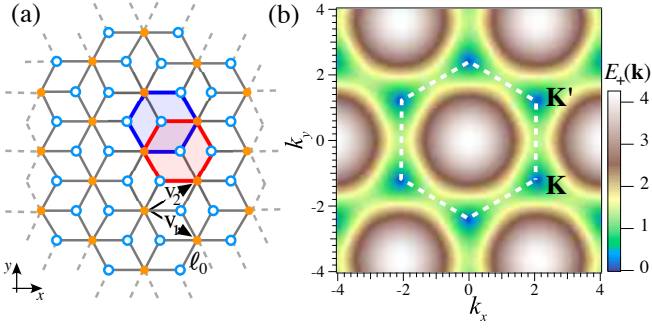}
\caption{\label{fig:one} (Color online). (a) The \ttt lattice.
It is characterized by translation vectors
$\mathbf{v}_1=(3/2;-\sqrt{3}/2)\ell_0$ and
$\mathbf{v}_2=(3/2;\sqrt{3}/2)\ell_0$, with lattice constant
$\ell_0$. Open circles mark the two sublattices A and B,
forming a HCL. Solid circles mark the hub sites H forming a
(larger) triangular lattice. (b) Contour plot of
$E_+(\mathbf{k})$, cf.\ Eq.~(\ref{eq:spectrum:b}). The dashed
hexagon defines the first Brillouin zone,
$\mathbf{K}=2\pi\ell_0^{-1}(1/3;-\sqrt{3}/9)$ and
$\mathbf{K}'=2\pi\ell_0^{-1}(1/3;\sqrt{3}/9)$ are two
non-equivalent Dirac points.}
\end{figure}

The recent preparation of single layers of
graphene~\cite{neto:2009} has attracted considerable attention,
since this solid state system displays quasi-relativistic motion
of electrons on a two-dimensional honeycomb lattice (HCL).
However, \emph{e.g.} due to disorder or impurities, many properties of real graphene cannot fully be
accounted for by the idealized Dirac-Weyl
Hamiltonian. In this Letter we present a detailed study of the
\ttt lattice~\cite{sutherland:1986} and show that cold fermionic atoms in
such an OL indeed behave as quasi-relativistic massless
Dirac-Weyl Fermions. Yet, the \ttt lattice
replaces the pseudo-spin $S=1/2$ of Dirac-Weyl particles in
the HCL by the larger value $S=1$. As one of its crucial
features, the \ttt lattice exhibits nodes with unequal
connectivity. The corresponding class of two-dimensional lattices, specifically
\emph{bipartite} lattices, has been studied extensively in the
past, with a particular focus on topological
localization~\cite{morita:1972,sutherland:1986}, frustration in
a magnetic field~\cite{vidal:1998,korshunov:2001},
and effects of spin-orbit
coupling~\cite{bercioux:2004}. The \ttt lattice,
illustrated in Fig.~\ref{fig:one}a, has a unit cell with three
different lattice sites, one six-fold coordinated site H, called
\emph{hub}, and two three-fold coordinated sites A and B, called
\emph{rims}. All nearest-neighbor pairs are formed by a rim
and a hub. The energy spectrum~\cite{sutherland:1986,vidal:1998} of the \ttt lattice
exhibits particle-hole symmetry and is characterized by three
branches. Two of them are linearly dispersing near the
$K$-points, in direct analogy to the energy dispersion known
from the HCL~\cite{neto:2009}. The third branch is
dispersionless, constantly equal to zero energy. These
localized states correspond to non-trivial solutions of the
Bloch equations for which the wave functions vanish identically on
the hub sites and have opposite amplitudes on the two different
rim sites. Localization giving rise to these dispersionless
states has a purely topological origin~\cite{morita:1972} and is
 quite robust against disorder~\cite{sutherland:1986}.

In this Letter we focus on low energy, long wavelength properties
of non-interacting Fermions in the \ttt lattice near half filling. As a main
result we find that in this parameter range the dynamics is governed by a Hamiltonian of the Dirac-Weyl form
%
%
%%%%%%%%%%%
\begin{equation}\label{eq:ham}
\mathcal{H} = v_\text{F}\, \mathbf{S}\cdot \left (\mathbf{p} - \frac{e}{c} \mathbf{A} \right).
\end{equation}
%%%%%%%%%%%
%
%
Here, $v_\text{F}$ is the Fermi velocity,
$\mathbf{p}=-\ii(\partial_x,\partial_y,0)$ the momentum operator
(we set $\hbar=1$) in the lattice $xy$-plane, and $\mathbf{A}$ the
vector potential associated with a magnetic field
$\mathbf{B}=(\partial_xA_y-\partial_yA_x)\mathbf{e}_z$
perpendicular to the plane. Equation~(\ref{eq:ham}) strikingly
resembles the Hamiltonian for Fermions in graphene.
However, as decisive difference, the pseudo-spin vector $\mathbf{S}=(S_x,S_y,S_z)$ now has total spin $S=1$
reflecting the three inequivalent lattice sites per
unit cell in \ttt, compared to two in the HCL. In the basis of
the eigenstates of $S_z$ the spin operators of (\ref{eq:ham})
can be expressed as $3\times 3$ matrices 

%
%
%%%%%%%%%%%
{\footnotesize
\begin{equation*}\label{eq:gm}
 S_x=\!\frac{1}{\sqrt{2}}
 \begin{pmatrix}
 0 & 1 & 0 \\
 1 & 0 & 1 \\
 0 & 1 & 0
 \end{pmatrix}\!\!,\
 S_y=\!\frac{1}{\sqrt{2}} \begin{pmatrix}
 0 & -\ii & 0 \\
 \ii & 0 & -\ii \\
 0 & \ii & 0
 \end{pmatrix}\!\!,\
 S_z=\!\begin{pmatrix}
 1 & 0 & 0 \\
 0 & 0 & 0 \\
 0 & 0 & -1
 \end{pmatrix}\!
\end{equation*}
}which satisfy angular momentum commutation relations.

%
%
%%%%%%%%%%%%%%
%%%%%%%%%%%%%%
%
%
For zero magnetic field the Hamiltonian (\ref{eq:ham}) is obtained  by
starting from the Schr\"odinger equation for the \ttt lattice in the
tight-binding approximation:

%
%
%%%%%%%%%%%%%
{\small \begin{subequations}\label{eq:bes}
\begin{align}
E \Psi_\text{H}(\mathbf{R}_\text{H}) & = -t \sum_{j} \Psi_\text{A}(\mathbf{R}_\text{H}+\tau_j)+\Psi_\text{B}(\mathbf{R}_\text{H}+\tau_{j+1}), \\
E \Psi_\alpha (\mathbf{R}_\alpha) & = -t \sum_j \Psi_\text{H}(\mathbf{R}_\alpha-\tau_j), ~~\alpha\in\{\text{A,B}\}.
\end{align}
\end{subequations}
%%%%%%%%%%%%
%
%
}Here, $\Psi_\alpha(\mathbf{R}_\alpha)$ is the amplitude of the
wave function on sublattice $\alpha=\text{A,H,B}$, and the $\tau_j$
connect nearest neighbors. Solving Eq.~(\ref{eq:bes}), we obtain the
energy spectrum
%
%
%%%%%%%%%%%
\begin{subequations}\label{eq:spectrum}
\begin{align}
E_0(\mathbf{k}) & = 0 \label{eq:spectrum:a}\\
E_\pm(\mathbf{k}) & = \pm t \big[ 6+4 \{ \cos[(\mathbf{v}_2-\mathbf{v}_1)\cdot \mathbf{k}] \nonumber \\ & \hspace{1cm} + \cos[ \mathbf{v}_1 \cdot \mathbf{k}] + \cos[ \mathbf{v}_2 \cdot \mathbf{k}] \}\big]^{1/2}\label{eq:spectrum:b}
\end{align}
\end{subequations}
%%%%%%%%%%%
%
%
where $E_\pm$ exhibit a linear dispersion, with Fermi velocity
$v_\text{F}=3t\ell_0/\sqrt{2}$~\cite{note:vf}, about either of the
Dirac-points $\mathbf{K}$ and $\mathbf{K}'$, as seen in Fig.~\ref{fig:one}(b). Next
we perform a long wavelength approximation about $\mathbf{K}$ 
%
%
%%%%%%%%%%%
\begin{equation}\label{eq:lwla}
\Psi_\alpha(\mathbf{R}_\alpha)= \ee^{\ii \mathbf{K}\cdot \mathbf{R}_\alpha} \psi_\alpha(\mathbf{R}_\alpha)
\end{equation}
%%%%%%%%%%%
%
%
by separating out the slowly varying part $\psi_\alpha$ of the amplitude.
Similarly, we can treat the neighborhood of $\mathbf{K}'$. In
the absence of short wavelength scattering processes, wave
functions from the vicinities of $\mathbf{K}$ and $\mathbf{K}'$
can be mapped onto each other by interchanging the two rim
components so that we focus on the vicinity of
$\mathbf{K}$ in the following. Inserting $\Psi_\alpha$ into
Eq.~(\ref{eq:bes}) finally yields the result~(\ref{eq:ham}) when arranging the
components $\alpha=\{\text{A,H,B}\}$ into a pseudo-spin triplet and
expanding to linear order in
$\mathbf{\kappa}=|\mathbf{k}-\mathbf{K}|$ for $\kappa\ll |\mathbf{K}|$.

Since $[\mathcal{H},\mathcal{J}_z]=0$, where $\mathcal{J}_z=L_z+S_z$ is the $z$-component of the total angular momentum and \mbox{$L_z=-\ii\partial_\varphi$}, we can choose simultaneous eigenstates of $\mathcal{H}$ and $\mathcal{J}_z$. At non-zero energy $E=\pm v_\text{F} \kappa$ these read in polar coordinates $(r,\varphi)$
%
%
%%%%%%%%%%%
\begin{equation}\label{eq:spinor:nomf}
\begin{pmatrix}
\psi_\text{A} \\ \psi_\text{H} \\ \psi_\text{B}
\end{pmatrix} = \mathcal{N} 
\begin{pmatrix}
 \xi_{-} J_{|m|-\xi_{-}}(\kappa r)\ee^{\ii (m-1)\varphi} \\
 \text{sgn}(E) \ii \sqrt{2} J_{|m|}(\kappa r) \ee^{\ii m\varphi} \\
 -\xi_{+} J_{|m|+\xi_{+}}(\kappa r)\ee^{\ii (m+1)\varphi}
\end{pmatrix}\;,
\end{equation}
%%%%%%%%%%%
%
%
with integer $m$, the eigenvalue of $J_z$.
Here, $J_m(x)$ is a Bessel function,
$\xi_{\pm}=\text{sgn}(m\pm0^+)$, where the infinitesimal $0^+$ ensures the
correct values for $m=0$, and $\mathcal{N}$ is a normalization constant. The
eigenstates at zero energy  are given by
%
%
%%%%%%%%%%%
\begin{equation}\label{zerotlevel}
\begin{pmatrix}
\psi_\text{A} \\ \psi_\text{H} \\ \psi_\text{B}
\end{pmatrix} = \tilde{\mathcal{N}}
\begin{pmatrix}
 \xi_- J_{|m|-\xi_{-}}(\kappa r)\ee^{\ii (m-1)\varphi} \\
 0 \\
 \xi_+ J_{|m|+\xi_{+}}(\kappa r)\ee^{\ii (m+1)\varphi}
\end{pmatrix}
\end{equation}
%%%%%%%%%%%
%
%
where $\tilde{\mathcal{N}}$ is again a normalization constant.
This solution has a finite amplitude only on the rim sites which
thus are \emph{topologically} disconnected. 
The zero energy solution (\ref{zerotlevel}) is infinitely degenerate with respect to the quantum numbers $\kappa$ and $m$.
The properties of this level, \emph{e.g.} the delta-like singularity in the density of states,
are distinctive features of the \ttt lattice and its
Dirac-Weyl behavior, as we shall see below.

Let us now consider the Hamiltonian (\ref{eq:ham}) in a
perpendicular, effective magnetic field. We use the symmetric gauge
$\mathbf{A}= B(-y,x,0)/2$. Defining the magnetic length
$\ell_B=\sqrt{2c/eB}$, the cyclotron frequency
$\omega_\text{c}=v_\text{F}/\ell_B$, and the
dimensionless radial coordinate $\rho =r/\ell_B$, 
the Schr\"odinger equation for eigenenergy $E=\omega_\text{c}\varepsilon$ reads
%
%
%%%%%%%%%%%
\begin{equation}\label{eq:sg:1}
\omega_\text{c}
\begin{pmatrix}
-\varepsilon & \mathcal{A} & 0 \\
\mathcal{A}^\dag & -\varepsilon & \mathcal{A} \\
0 & \mathcal{A}^\dag & -\varepsilon
\end{pmatrix}
\;
\begin{pmatrix}
\psi_{\text{A}} \\ \psi_\text{H} \\ \psi_{\text{B}}
\end{pmatrix}
= 0\,.
\end{equation}
%%%%%%%%%%%
%
%
Here
$\mathcal{A}=\frac{\ii}{\sqrt{2}} \ee^{-\ii \varphi}\left(-\partial_{\rho}+\rho-L_z/\rho\right)$.
The set of first order differential
Eqs.~(\ref{eq:sg:1}) can be recast into second-order
differential equations. For the hub component
$\psi_\text{H}(\rho,\varphi)~\propto~\phi_\text{H}(\rho) \ee^{\ii m
\varphi}$ we get
%
%
%%%%%%%%%%%
\begin{equation}\label{eq:phih:mf}
\phi_\text{H}''(\rho)+\frac{\phi_\text{H}'(\rho)}{\rho}+\left( 2m-\rho^2-\frac{m^2}{\rho^2}+\varepsilon^2\right)\phi_\text{H}(\rho)=0
\end{equation}
%%%%%%%%%%%
%
%
and similar equations hold for $\phi_\text{A/B}(\rho)$. The general
solution at $\varepsilon\neq 0$ is given by
%
%
%%%%%%%%%%%
\begin{equation}\label{eq:sol:mf:1}
 \begin{pmatrix}
 \psi_\text{A} \\ \psi_\text{H} \\ \psi_\text{B}
 \end{pmatrix}
 =
 \begin{pmatrix}
 \mathcal{N}_\text{A} \;\rho^{|m|-\xi_{-}} L_{n+\eta_{-}}^{|m|-\xi_{-}}(\rho^2) \ee^{\ii (m-1)\varphi} \\
 \mathcal{N}_\text{H} \,\text{sgn}(E)\; \rho^{|m|} L_n^{|m|}(\rho^2) \ee^{\ii m\varphi}\\
 \mathcal{N}_\text{B}\; \, \rho^{|m|+\xi_{+}} L_{n-\eta_{+}}^{|m|+\xi_{+}}(\rho^2) \ee^{\ii (m+1)\varphi}
 \end{pmatrix}\ee^{-\frac{\rho^2}{2}}\!\!
\end{equation}
%%%%%%%%%%%
%
%
with integer $m$ and positive integer $n$. 
The $\mathcal{N}_\alpha$ are appropriate numerical factors~\cite{note:2}. Here,  $L_a^b(x)$ is
an associated Laguerre polynomial of order $a\ge 0$, and
$\eta_\pm=\theta(m\pm0^+)$, where $\theta$ is the unit-step
function. The eigenenergies of the states (\ref{eq:sol:mf:1}), the \emph{Landau levels}, are
found to read
%
%
%%%%%%%%%%%%%%%%
\begin{equation}\label{llevels}
 E_{n,m} = \pm  2\omega_\text{c} \sqrt{n+|m|\theta(-m)+\frac{1}{2}}\,.
\end{equation}
%%%%%%%%%%%%%%%%%
%
%
They scale with the square root of the index $n$ and $m$ and display a finite energy value even for $n=m=0$.
These peculiarities are in analogy with the unusual Landau levels of the HCL
that have received broad attention in the context of graphene~\cite{graphenelandaulevels,demartino}.
 
Equation (\ref{eq:sg:1}) deserves special care  at zero energy. Then, two solutions are allowed: a zero-energy Landau level and a topological solution. The former is characterized by a single non-vanishing component 
%
%
%%%%%%%%%%%
\begin{equation}\label{zero:ll}
\psi_\text{A}\propto \ee^{-\frac{\rho^2}{2}} \rho^{m-1} \ee^{\ii (m-1)\varphi}
\end{equation} 
%%%%%%%%%%%%%%%
%
%
with $m\ge 1$~\cite{note:3}. The other solution has a vanishing hub component and reads
%
%
%%%%%%%%%%%%%%
\begin{equation}\label{eq:sol:mf:2}
 \begin{pmatrix}
 \psi_\text{A} \\ \psi_\text{H} \\ \psi_\text{B}
 \end{pmatrix}\!\!
 =\!\!
 \begin{pmatrix}
    \tilde{{\cal N}}_\text{A}\;\rho^{|m|-\xi_{-}} L_{n+\eta_{-}}^{|m|-\xi_{-}}(\rho^2)\ee^{\ii (m-1)\varphi} \\
    0  \\
    \tilde{{\cal N}}_\text{B}\;\rho^{|m|+\xi_{+}} L_{n-\eta_{+}}^{|m|+\xi_{+}}(\rho^2)\ee^{\ii (m+1)\varphi} \end{pmatrix}\ee^{-\frac{\rho^2}{2}}\!\!\!
\end{equation}
%%%%%%%%%%%%%%%
%
%
with $n\ge1$~\cite{note:4}. 
The zero-energy Landau level (\ref{zero:ll}) has the same degeneracy as the other Landau levels
at finite energies (\ref{llevels}). In the case of the HCL a similar solution is responsible
for the atypical quantum-Hall effect observed in transport
experiments through graphene layers~\cite{grapheneqhe}. The Atiyah-Singer index theorem~\cite{indexthyrefs} connects these zero-energy Landau levels with the structure of the Dirac-Weyl Hamiltonian. Thus,
contrary to the finite energy Landau levels, they are robust against fluctuations of the electrostatic and magnetic fields. 
On the other hand, the existence of the topological solution~(\ref{eq:sol:mf:2}) is due to the bipartite structure of the \ttt lattice. The two types of zero-energy solutions cannot be distinguished on the mesoscopic scale.
Their level degeneracies just add up and enhance the total degeneracy at zero energy.
This feature should be observable in quantum Hall effect experiments.

\begin{figure}[!t]
\centering
\includegraphics[width=0.9\columnwidth]{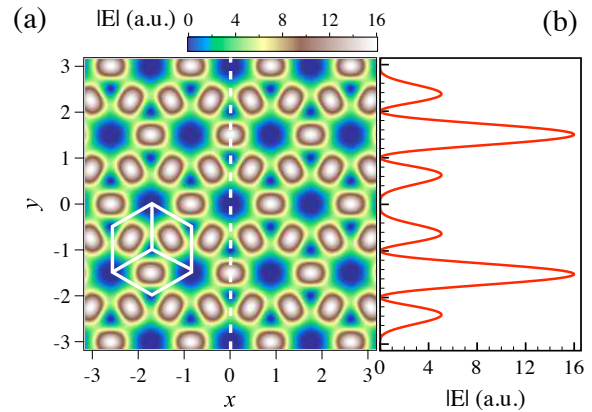}
\caption{\label{figure:2} (Color online) (a) Distribution of the
laser field intensity for generating a \ttt lattice. (b) Cut of
the field intensity along the $y$-axis.}
\end{figure}

%
% TAU_3 LATTICE IN AN OPTICAL LATTICE
%
Experimentally, the \ttt structure can be realized as an OL created by  three counter propagating
pairs of laser beams with the same wavelength $\lambda =
3/2\ell_0$ which divide the plane into six sectors of
$60^\circ$. All six laser beams are linearly polarized with the
electrical field in the \mbox{$xy$-plane}.
Orienting the polarization
of one pair of lasers along the $y$ axis, $\mathbf{E}_1 = (0 , E_y,0)$,
the other two pairs are obtained by rotating $\mathbf{E}$ by
$120^\circ$ around the $z$-axis.
This arrangement produces the profile of laser intensity depicted in Fig.~\ref{figure:2}(a); Fig.~\ref{figure:2}(b)
shows a cut of the field intensity along the $y$-axis.
It is apparent that the  hopping probability between rim
sites is exponentially small compared to the hopping probability
between neighboring hub-rim pairs. Contrary to
other theoretical proposals for the observation of
massless Fermions in OLs, we do not  require  the presence of  
time-dependent potentials~\cite{lim:2008}, staggered gauge fields~\cite{hou:2009}, external gauge fields~\cite{goldman:2009}, and the proposed laser configuration does not create spurious lattice minima~\cite{zhu:2007}.

%
%
%%%%%%%%%%%%
%%%%%%%%%%%%
%
%
The tight-binding Hamiltonian and its long wave length
approximation (\ref{eq:ham}) are valid for a \ttt lattice
populated by single-component fermionic atoms, \emph{e.g.}
$^{40}$K or $^6$Li. Indeed, for single-component Fermions, the atomic
collisions are negligible at low
temperature~\cite{Bloch:2008p894}. From the experimental point
of view, time-of-flight imaging via light
absorption~\cite{koehl:2005} can be used in order to detect the
presence of massless Dirac-Weyl Fermions. The harmonic trap potential
$V(\mathbf r)=m\omega^2\mathbf{r}^2/2$ confining the fermionic
cold atom gas is ramped down slowly enough for the atoms to stay
adiabatically in the lowest band while their quasi-momentum is
approximatively conserved. Under these conditions, free Fermions
expand with
ballistic motion and, from the measured absorption images, it is
possible~\cite{ho:2000,umucalilar:2008} to reconstruct the initial reciprocal-space density profile
of the trapped gas. Then, the local
density approximation is typically well satisfied and the
local chemical potential can be assumed to vary with the radial
coordinate as $\mu(\mathbf{r}) = \mu_0 -V(\mathbf{r})$, where $\mu_0$ is the
chemical potential at the center of the trap.
For a system of cold atoms at temperature $T$, the atomic density is uniquely
determined by the chemical potential
%
%
%%%%%%%%%%%%
\begin{equation}\label{eq:twoelve}
n(\mu) = \frac{1}{\mathcal{S}_0} \int f(\mathbf{k},\mu) \, d \mathbf{k} \,.
\end{equation}
%%%%%%%%%%%%
%
%
Here $\mathcal{S}_0$ is the area of the
first Brillouin zone of the \ttt lattice, and
$f(\mathbf{k},\mu)=[\exp[(E_\alpha(\mathbf{k})-\mu)/k_\text{B}T]+1]^{-1}$
is the Fermi distribution function, where $E_\alpha(\mathbf{k})$
is the energy spectrum of the \ttt lattice,
cf.\ Eq.~(\ref{eq:spectrum}).
%
%
%%%%%%%%%%%%
\begin{figure}[!t]
\centering
\includegraphics[width=0.7\columnwidth]{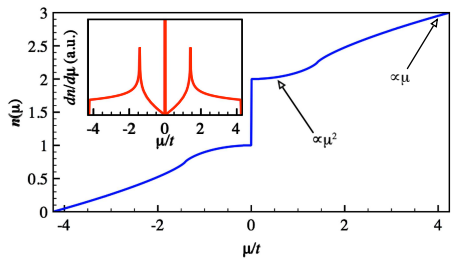}
\caption{\label{figure:3} (Color online) The number of atoms $n$ per unit cell at zero
temperature as a function of $\mu/t$, ignoring physical spin and
valley degeneracies, where $t$ is the nearest neighbor hopping
amplitude, cf.\ Eq.~(\ref{eq:bes}). Inset: density of states,
$dn/d\mu$, versus $\mu/t$.}
\end{figure}
%%%%%%%%%%%%
%
%
Figure~\ref{figure:3} shows the atomic density $n$
as a function of the chemical potential $\mu$.   
The contribution from the highly degenerate topological band (\ref{zerotlevel}) manifests itself at $\mu=0$ as a sharp jump in the atomic density. This feature is specific to the \ttt lattice. 
For small finite $\mu$ we see that $n$ increases (decreases) proportional to $\mu^2$, which 
reflects the linear dispersion of massless Fermions near the band center as well as particle-hole symmetry~\cite{onsite}. 
On the
contrary, for values of $\mu$ close to the maximum or minimum of
the energy band, \emph{i.e.}\ far away from the band center, where the
long wave length approximation can no longer be applied,
$n$ varies proportional to $\mu$.

In conclusion, we have discussed an experimental setup for the observation of Dirac-Weyl Fermions in a \ttt OL.
In particular, we have shown that in the low energy and long wavelength approximation the \ttt Hamiltonian describes massless Dirac-Weyl Fermions.  This generalizes results known for the case of the HCL, however, with pseudo-spin $S=1$. The Dirac cones manifest themselves in a quadratic dependance of the atomic density  on the chemical potential. Besides the Dirac-Weyl Fermions, the \ttt energy spectrum features a dispersionless branch  of localized states. These states give rise to a pronounced finite jump in the atomic density, a hallmark of the \ttt lattice. We have also studied the effects of a effective magnetic field perpendicular to the lattice. This leads to the formation of unusual Landau levels resembling those known for the case of the HCL. In particular a zero-energy Landau level coexists with the dispersionless level of localized states.

%
% ACKNOWLEDGMENT
%
We gratefully acknowledge helpful discussions with
A.~De~Martino, R.~Egger, T.~Esslinger, P.~H\"anggi, and M.~Rizzi.
This work was supported by the Excellence Initiative of the German
Federal and State Governments.

%\break


\begin{thebibliography}{99}
\bibitem{note:vf} Note that $v_\text{F}$ differs by a factor $\sqrt{2}$ from the Fermi velocity of a HCL with the same parameters $t$ and $\ell_0$.
\bibitem{note:3} For $m<1$ the solution of the associated differential equation is not regular for $\rho\to0$.
\bibitem{onsite} Actually, a small difference in the on-site energies,
$\varepsilon_{\rm H}$ and $\varepsilon_{\rm R}$ for hub and rim
sites, respectively, will cause a gap in the dispersing part of
the energy spectrum, though leaving the topological band untouched.
This then modifies $n(\mu)$ in very close vicinity $\sim |\varepsilon_\text{H}-\varepsilon_\text{R}|$ of the
sharp jump.
\end{thebibliography}
\end{document}